# Atomistic simulation of the Coupled adsorption and unfolding of protein GB1 on the polystyrenes nanoparticle surface

*Huifang Xiao[1,†], Bin Huang[1,†], Ge Yao[1], Wenbin Kang[1,3], Sheng Gong[2], Hai Pan[1], Yi Cao, Jun Wang[1], Jian Zhang\*[1], Wei Wang\*[1]*

[1]School of Physics, Collaborative Innovation Center of Advanced Microstructures, and National Laboratory of Solid State Microstructures, Nanjing University, Nanjing, China

[2]Department of Pharmaceutics, Jinling Hospital, Nanjing University School of Medicine, Nanjing, China

[3]School of Public Health and Management, Hubei University of Medicine，Hubei University of Medicine, Shiyan, China

[†] These authors contribute equally

**Corresponding authors**: *jzhang@nju.edu.cn (JZ), *wangwei@nju.edu.cn (WW)

**ABSTRACT**

Protein adsorption/desorption upon nanoparticle surfaces is an important process to understand for developing new nanotechnology involving biomaterials, while atomistic picture of the process and its coupling with protein conformational change is lacking. Here we report our study on the adsorption of protein GB1 upon a polystyrene nanoparticle surface using atomistic molecular dynamic simulations. Enabled by metadynamics, we explored the relevant phase space and identified three protein states; each protein state involved both the adsorbed and desorbed states. We also studied the change of secondary and tertiary structures of GB1 during adsorption, and the dominant interactions between protein and surface in different adsorbing stages. From the simulation results we obtained a scenario that is more rational and complete than the conventional one. We believe the new scenario is more appropriate as a theoretical model in understanding and explaining experimental signals.

**Introduction**

With the rapid development of nanotechnology, more and more attention has been paid to the combination of nanomaterials with biomaterials to make novel functional materials or tiny devices for drug delivery, bioimaging, sensing, diagnosing, or more speculatively, nano-enzymes and nanorobots[1-17]. The researches involve the interaction of nano-scaled surfaces with biomaterials, especially proteins. Therefore, it is essential to study protein adsorption/desorption upon various surfaces of nanomaterials. However, experimental techniques for studying the structures or dynamics of the protein-surface interactions usually lack atomistic resolution. Molecular simulation can alleviate this issue and have been widely used to confirm or complement experimental

results[18-33].

Molecular simulation of the protein adsorption at the atomistic level is by no means a trivial problem, since in addition to the protein folding/unfolding barriers, the simulation needs to overcome the free energy barriers associated with the adsorption/desorption processes, which are often comparable to the former[28,34]. Although there are a large volume of atomistic simulations on this subject[23-33], most of them cannot reach the time scale of experiments due to the difficulty of overcoming these barriers and the inherent ruggedness of the free energy landscape.

In this work, we use a well-characterized protein GB1, the B1 domain of Streptococcal protein G, and the polystyrene (PS) nanoparticle surface as our modeling system to study the atomistic picture of the adsorption process and the coupled protein conformational change. The thermodynamics and kinetics of this system have been studied in detail in our lab by using a stopped-flow fast mixing technique[34]. It was suggested that there are three major states, the folded but desorbed state, the folded and adsorbed state, and the unfolded and adsorbed state; the kinetics can be described by a fast adsorption followed by a slow reversible unfolding of GB1; all rate constants were measured and the free energy profile was constructed. However, the experiment did not identify which parts of the protein attach to the surface and lacked atomistic information of their interaction. Here we attack this problem with all-atom molecular dynamic simulations. To overcome the barrier-crossing problem, we employed metadynamics[35-38], which periodically modified the effective energy by adding small repulsive Gaussian potentials and thus enforced escaping from local minima. This technique was recently used to exhaustively sample a peptide adsorption on two self-assembled monolayer (SAM) surfaces[28]. From the computational data of metadynamics, we constructed free energy landscape, identified various adsorption states, and analyzed the corresponding protein structures. From the above results we obtained a scenario that is more complete than the experimental one. We believe the new scenario better to describe the protein adsorption upon the surface of nanoparticles.

**METHODS**

**Modeling of the PS nanoparticle surface**

A total of 96 polystyrenes (PS) of 10-monomer were stacked into 4 layers to mimic the surface of LATEX nanoparticles in the experiment[34]. Considering the large size of the nanoparticles in the experiment, the curvature effect of the nanoparticles was neglected in this study. To mimic the effect of the electrical double Layer associated with the nanoparticle, 18 Cl- anions were randomly put on the top of the PS surface and restrained along the z-axis, which was perpendicular to the PS surface. The ions were allowed to diffuse freely on the two dimensional PS surface. The restraints were necessary to mimic a stable surface charge. Otherwise, the fluctuation of the surface charge would be very large since the absolute number of ions was small. The number of the restrained ions was calculated from the Poisson-Boltzmann equation[39] and described in detail in the next section.

The PS surface was first subjected to an energy minimization of 2000 steps, followed by a molecular dynamics (MD) simulation of length 500ps at 300 K, with all α-carbon atoms restrained. After that, we released the restraints on the top three layers while kept that on the bottom layer, and then run a MD simulation of length 1ns at 300K. This resulted in a relaxed PS surface. The top three layers were found to be adsorbed to the bottom layer automatically. The last frame of the simulation was used to construct the protein-surface system described in the following section.

**Calculation of the surface charge on the nanoparticle**

Around a charged colloidal particle immersed in an electrolyte solution, counter ions tend to approach the particle surface and neutralize the particle surface charges, forming an ionic cloud. The ionic cloud together with the particle surface charge forms an electrical double layer[39]. The potential distribution around charged colloidal particles plays a fundamental role in their interfacial electric phenomena.

The electric potential $\psi(x)$ at position x outside the particle, when the potential is low, can be described by the linearized Poisson-Boltzmann equation[39],

$$\frac{d^2\psi}{dx^2} = \kappa^2\psi, \qquad (1)$$

where $\kappa$ is the Debye-Hückel parameter and $\kappa = (\frac{1}{\varepsilon_0\varepsilon_r kT}\sum_{i=1}^{N} z^2 q^{-2} n_i^\infty)^{\frac{1}{2}}$, where $n_i^\infty$ is the bulk concentration (number density) of the *i*-th ionic species with valance $z_i$, $q$ is the electron charge, $k$ is the Boltzmann constant and $T$ is the temperature. The solution of the equation is $\psi(x) = \psi_0 e^{-\kappa x}$, where $\psi_0 = \frac{\sigma}{\varepsilon_0\varepsilon_r\kappa}$, $\sigma$ is the charge density of the particle surface. The zeta-potential is defined as $\zeta = \psi\left(\frac{1}{\kappa}\right) = \frac{\psi_0}{e}$, which can be measured experimentally. Therefore, we have $\sigma = \varepsilon_0\varepsilon_r\zeta\kappa e$.

For Nacl, we have $\kappa = (\frac{2q^{-2}n^\infty}{\varepsilon_0\varepsilon_r kT})^{\frac{1}{2}}$ and thus $\sigma = \varepsilon_0\varepsilon_r\zeta e(\frac{2q^{-2}n^\infty}{\varepsilon_0\varepsilon_r kT})^{\frac{1}{2}}$. According to the experiment[34], $n^\infty = 137nM$, $\zeta = -30.2mV$, $T = 298K$, and $\varepsilon_{rNacl} = 75$. Therefore, we have $\sigma = \varepsilon_0 \varepsilon_r \zeta e (\frac{2q^{-2}n^\infty}{\varepsilon_0 \varepsilon_r kT})^{\frac{1}{2}} = 0.4235 e^-/nm^2$.

The surface area of the particle surface in the simulation is estimated to be $S = 42 nm^2$, therefore, the number of negative charges on the PS surface is $N = \sigma S \approx 18$.

**System setup**

A protein GB1, the B1 domain of Streptococcal protein G (pdb code: 3GB1), was put at a distance of ~2.2 nm from the top layer of the PS surface, as shown in Fig. 1. At this distance it had not direct contacts with the surface. A cubic box of TIP3P water was added to solvate the protein and the PS surface. The resulted dimension of the box was 7.9 x 8.6 x 9.9 nm. A total of 61 Na+ and 38 Cl- ions were added to achieve a salt concentration of 137mM[34]. Periodic boundary conditions were applied in all three dimensions. The PS surface was along the x-y plane. The α-carbon atoms of all the polystyrenes were restrained. A repelling potential was added near the top of the box to prevent the protein drifting close to the bottom layer of the upper image of the PS surface. Atomic charges and atom types for the PS were assigned by antechamber[40]. Force filed parameters for all of the bonds, angles, dihedrals were taken from AMBER99SB-ILDN[41], which had been shown to exhibit considerably better agreement with the NMR data. The electrostatic interaction was treated using PME with a cutoff of 1.0nm. The same cutoff was used in the calculation of the van der Waals interactions. All bonds were constrained using the LINCS algorithm and the MD time step was set to 2fs. Berendsen algorithm was used for the temperature coupling. All simulations were performed with GROMACS (v4.6.7)[42-43].

**Simulation procedures**

The whole system was first subjected to a steepest descent minimization of 1000ps with all heavy atoms restrained, followed by a similar minimization of 1000ps without restraints. A MD simulation in NPT ensemble (300 K, 1 bar, 2ns) and a MD simulation in NVT ensemble (300 K, 2ns) were then carried out successively to further relax the system and prepare system for the simulations that follow.

To overcome the folding/unfolding barriers and the even stronger adsorption/desorption barriers, we adopted metadynamics[35-38], which added bias potentials periodically along a set of pre-chosen Collective Variables (CVs) to help the system escape basins of attraction. In metadynamics, the overall external Gaussian potentials added to the system at time $t$ is given by[35-38]

$$V(S(x), t) = \omega \sum_{t'=\tau, 2\tau, \cdots, t'<t} exp\left(\frac{(S(x)-s(t'))^2}{2\delta s^2}\right),$$

where $x$ is the system configuration, $s(t) = S(x(t))$ is the value taken by the CVs at time t, $\omega$ is the Gaussian height, $\delta s$ the Gaussian width, and $\tau$ determines the frequency of adding Gaussian potentials. The basic assumption of metadynamics is that $V(S(x), t)$ after a sufficiently long time provides an estimate of the underlying free energy,

$$\lim_{t \to \infty} V(s, t) \sim -F(s).$$

For the adsorption process studied here, the bias was applied upon the distance between protein and surface. Specifically, the distance was calculated as the center of mass of the protein GB1 and that of the top layer of the surface.

The Metadynamics simulations were carried out at 300K in a NVT ensemble. The height of the Gaussian potentials was set to 0.15 kJ/mol and their width was 0.1 nm. The deposition rate of the Gaussian potentials was 1000 ps$^{-1}$. Note that the height and deposition rate of the Gaussian potentials were significantly smaller than that usually reported in literatures. This is to avoid the artifacts associated with large Gaussian height and deposition rate as much as possible. In total four metadynamics simulations were performed, each starting from a different protein orientation with respect to the PS surface. Each simulation lasted for 220 nanoseconds. The simulations were stopped after sufficient number of back and forth sampling of the CV space were observed.

All metadynamics simulations were performed with GROMACS (v4.6.7)[42-43] and the PLUMED(v2.2.0) plug-in[37-38].

**Data analysis**

Two-dimensional free energy landscapes (FELs) were calculated from the data of metadynamics using a reweighting technique[44]. Note that in the simulations only one bias was applied, which was on the distance between protein and surface. The two-dimensional FELs were obtained by a reweighting technique. The two CVs on which the FELs were projected were the distance between protein and surface, and the Root Mean Square Distance (RMSD) of the protein with respect to its native structure. A cluster analysis was carried out to examine the structures of the detected basins of attraction. Specifically, the cluster algorithm counted the number of neighbors for each structure using RMSD cut-off, took the structure with the largest number of neighbors with all its neighbors as a cluster and then eliminated it from the pool of structures; the algorithm repeated this procedure for the remaining structures until none was left. The algorithm was implemented in the Gromacs software[45]. The change of the secondary structures as a function of simulation time was analyzed with DSSP[46-47].

**Results**

Figure 2 gives the free energy landscapes (FELs) calculated from metadynamics simulations. All four FELs consistently show three major states, corresponding to the folded state of the protein (RMSD below 0.2nm), the intermediate state (RMSD roughly from 0.25nm to 0.5nm), and the unfolded state (RMSD>0.5nm), respectively. On the FELs we superimpose the corresponding trajectories to show their time evolution. The four FELs, together with the superimposed trajectories, give a consistent picture described as follows. 1) In general, the trajectories move from low RMSD regions to high ones with time and pass through the F-, I-, and U-states in turn. And in each state, the protein transits between the desorbed and absorbed states many times. 2) At the early stage, the protein fluctuates between the adsorbed and desorbed status while mostly stays at the F-state, suggested by the small RMSDs. 3) Then the protein partially unfolds and enters the intermediate state, indicated by the red arrows. Note that these unfolding events are only observed at small protein-surface distances, tentatively attributed to the denaturing effect of the surface. 4) In the intermediate state, the protein fluctuates between the adsorbed and desorbed states many times before it further unfolds and enters the unfolded state, indicated by the magenta arrows. Again, these unfolding events are only observed when the protein is near the surface. 5) Spontaneous refolding events from the intermediate states back to the native states are also observed, shown by the yellow arrows in Fig. 2(a) and 2(d). Interestingly, these events happen at large protein-surface distance, where the denaturing effect of the surface upon protein is minimal.

Caution should be given regarding the barrier heights and transition frequencies between the states. Fig. 2 seems to imply that the barriers between F-, I-, and U-states are significantly larger than those between the adsorbed and desorbed states. However, the feature may be due to the artifact of metadynamics, which enhanced the sampling along the protein-surface distance while not along RMSD. Therefore, it is only safe to compare the barrier heights and transition frequencies of the reactions that occur along the same CV, either the distance or the RMSD. We do not compare the reactions that happen along different CVs. Furthermore, the absolute values of the barrier heights cannot be compared to the experiments either. This is because metadynamics is a non-equilibrium algorithm in nature, and the calculated barrier heights are somehow dependent on the parameters, particularly the depositing rate of the Gaussian potentials.

The evolution of the protein secondary structures as a function of time is shown in Fig. 3. Each figure corresponds to a trajectory in Fig. 2. All the figures show a consistent scenario. That is, under the affection of the surface, the helical region (A23 to D36) breaks first, roughly starting at 40ns and finishing at 90-120ns. In comparison, the β-contents are much more stable. Here we label the four β-strands of the protein with S1, S2, S3, and S4, respectively, in the order of the sequence from the N-terminus to the C-terminus. According to the first and third trajectories, the second hairpin (formed by S3 and S4 and denoted as S3-S4 hereafter) breaks at 90-120ns, while the first hairpin (S1-S2) holds until the end of the trajectories. In contrast, the second trajectory shows both hairpins are stable until 210ns. Interestingly, the fourth trajectory indicates an early unfolding of the second hairpin and a refolding back later. In general, the β-contents are less affected by the adsorption compared with the helix, possibly due to their flat geometries, which are more compatible with the geometry of the surface.

To further understand the nature of the interactions between protein and surface, we analyzed the FELs and trajectories further. Here we present the results for the first trajectory that in Fig. 2(a) and omit that for the other three, since they show very similar behaviors. For each of the three states in Fig. 2(a), we first collected the conformations within the state based on their RMSDs, plus the condition that the distance between protein and the surface was less than 2.2nm. The latter condition was applied because we were only interested in the adsorbed structures. From the collected conformation we then calculated the average vdW and electrostatic energies between protein and surface, and mapped the energies onto residues. The results are shown in Fig. 4. We also performed cluster analysis for the conformations for structural investigation. The results are given in Fig. 5.

The energy and structure analyses for the folded state can be seen in Fig. 4(a)-(b) and Fig. 5(a), respectively. This state corresponds to the stage when the protein first interacts with the surface. The calculation was based on 18114 conformations collected from the folded state. It can be seen that the vdW interaction between protein and surface is the leading force for the adsorption, to which both hydrophobic and hydrophilic residues contribute. This aspect is interesting considering that the PS surface is hydrophobic in nature. The central structure of the largest cluster given in Fig. 5(a) represents 13% of all the collected conformations. The RMSD of this structure with respect to the native structure is 0.11nm, and the distance between the center of mass of the protein and the surface is 1.5nm. The "hot spot" residues, i.e., that have large contributions to the interaction, include E19 and V21 in the loop, the residues from D22 to Q32 in the helix, and from E42 to A48 in the β-strand S3. The electrostatic interaction also contributes for the adsorption, mostly via the charged residue K28.

The energy and structure analyses for the intermediate state are shown in Fig. 4(c)-(d) and Fig. 5(b)-(c), respectively. The calculation was based on 7498 conformations collected from this state. It can be seen that the main adsorption force is still the vdW interaction. The hot spot residues include M1 in the N-terminus, V21 in the loop, D22, A23, A24, F30, K31 and N35 in the helix, W43, Y45, and D47 in the β-strand S3, and F52 in the strand S4. About sixty percent of them are hydrophobic residues. As for the electrostatic interaction, the contribution from K28 seen in the previous case vanishes while that from M1 and K50 appears. Compared with the previous case, the vdW interactions between protein and surface become much stronger, consistent with the largely deformed structures shown in Fig. 4(c) and (d). The two structures in Fig. 5(b)-(c) represent 59% and 13% of all collected conformations, and their RMSDs with respected to the native structure are 0.37 and 0.34nm, respectively. It can be seen that the overall structure is more open than the native one, having the inner hydrophobic residues partially exposed and attached to the surface. The helix is partially unfolded while the β-sheet is almost intact. This feature is consistent with the secondary structure evolution given in Fig. 3, which shows that the stabilities of the helix, the second hairpin, the first hairpin decrease in order.

The results for the unfolded state are given in Fig. 4(e)-(f) and Fig. 5(d)-(f). The calculation was based on 35855 conformations collected in this state. Compared with the previous two cases, more residues have contributions to the adsorption, including that in the N-terminus (M1, T2, Y3, and L5), in the loop regions (A20, V21, V39, and D40), in the helix (D22, A23, A24, and F30) and in the second β-hairpin (T18, W43, Y45, F52). About seventy percent of them are hydrophobic residues. The magnitudes of the vdW interactions are even larger than in the previous cases. As for the electrostatic interactions, M1, K28, and K50 make significant contributions. The central structures of the largest three clusters shown in Fig. 5(d)-(f) represent 21%, 8%, and 7% of all the collected conformations, respectively. It can be seen that the tertiary contacts between helix and β-sheets are completely lost, and the protein is essentially flat and lying on the surface. The helix is partially unfolded while the β-contents hold to some extent, consistent with the secondary structure analysis given in Fig. 3.

**Discussion and Conclusion**

In comparison with the experiment[34], our simulations support its conclusion that the vdW and electrostatic interactions play important roles in the adsorption process. In addition, our simulations reveal more adsorption states and their structural and energetic details. Analysis of these states shows an adsorption process summarized as follows. The protein usually attaches to the surface via the N-terminus of the helix and the residues in the loop and β-sheet that are close to the helix. As the protein progressively unfolds due to the denaturing effect of the surface, the "hot spots" gradually spread to the other regions. At last, the protein becomes mostly flat and is adsorbed upon the surface via the unfolded helix and one side of its β-sheet. Along with the adsorption and unfolding process, the percentage of hydrophobic residues that contribute to the vdW interaction between protein and surface progressively increases. This reflects the universal property that the protein surface residues are mostly hydrophilic while the inner ones are mostly hydrophobic. As a result, the surface residues are more relevant in the early adsorption stage while the inner ones are more relevant in the later stages, when the protein becomes more open and exposes its inner residues.

The experiment suggested that the kinetics can be described by a fast adsorption of GB1 upon the surface followed by a slow reversible unfolding[34], which is a "sequential scenario" as shown in Fig. 6(a). However, the simulations show that the scenario is more complicate, which we refer to as a "network" model and depict it in Fig. 6(b). In this new scenario, the protein may be adsorbed on the surface and unfolds afterwards, similar to the sequential model; or it may detach at any stage from the surface and transform to other conformational states; it may also be adsorbed back to surface again. The protein undergoes frequent transitions between the six states in Fig. 6(b). In the new model there is no apparent event sequence. The transition frequencies of the events need to be discussed. According to the simulations, the unfolding events were mostly observed in the adsorbed states, while seldom in the desorbed states. In comparison, the folding events (from I to F) were only observed in the desorbed state. In short, the probabilities for unfolding in the desorbed state and for folding in the absorbed state are low. This can be understood as follows. The PS surface is hydrophobic in nature and hence provides a denaturing environment for the protein, while the solvent favors the native state at the present simulation setup.

Overall, the new scenario is significantly different from the previous one, and we believe it is more complete and appropriate for describing the adsorption of GB1 on the PS surface. It may also reflect a general mechanism of protein adsorption on the surface of nanoparticles. Furthermore, it is interesting to see if the adoption of the new scenario as a theoretical model in fitting experimental signals would give different outcomes.

Caution should be given regarding the potential flaw of the AMBER99SB-ILDN force field used in the simulations. The forced field modified the side-chain torsion parameters of four residues that appear to be most problematic in ff99SB when comparing the rotamer distribution observed in MD simulations with that in the PDB database. The new parameters were obtained by fitting to new QM data and validated with microsecond-timescale MD simulations[41]. It has been shown to perform well by many works[48-51]. However, it was also reported to tend to increase helical content[52], encourage global contacts[53], give stronger interaction of ARG and LYS with the lipid phosphate groups, or generally overestimate the potential energy of protein-protein interactions at the expense of water-water and water-protein interactions[54-55]. The results presented here may be affected by these potential flaws.

**Supporting Information**

A movie showing the adsorption/desorption and unfolding of proteins GB1 on the PS surface is given.

**Declaration**

The authors have declared no conflict of interest.

ACKNOWLEDGMENT

The authors acknowledge the National Natural Science Foundation of China (No. 11274157, 31671026, 11334004), the National Basic Research and Development Program of China (2013CB834100), and Priority Academic Program Development (PAPD) project of Jiangsu Higher Education Institutions. The authors also acknowledge HPCC of Nanjing University and Shenzhen Supercomputer Center for the computational support.

**References**

1. Sarikaya M, Tamerler C, Jen A K Y, Schulten K, Baneyx F. Molecular biomimetics: nanotechnology through biology. Nature materials 2003, 2: 577-585.

2. Katz E, Willner I. Integrated nanoparticle-biomolecule hybrid systems: synthesis, properties, and applications. Angew Chem Int Ed Engl 2004, 43: 6042-6108.

3. Astier Y, Bayley H, Howorka S. Protein components for nanodevices. Curr Opin Chem Biol 2005, 9: 576-584.

4. Ying E, Li D, Guo S, Dong S, Wang J. Synthesis and bio-imaging application of highly luminescent mercaptosuccinic acid-coated CdTe nanocrystals. PLoS One 2008, 3: e2222.

5. Nel A E, Madler L, Velegol D, Xia T, Hoek E M V, Somasundaran P, Klaessig F, Castranova V, Thompson M. Understanding biophysicochemical interactions at the nano-bio interface. Nature materials 2009, 8: 543-557.

6. Petros R A, DeSimone J M. Strategies in the design of nanoparticles for therapeutic applications. Nat Rev Drug Discov 2010, 9: 615-627.

7. Mahmoudi M, Lynch I, Ejtehadi M R, Monopoli M P, Bombelli F B, Laurent S. Protein-nanoparticle interactions: opportunities and challenges. Chem Rev 2011, 111:5610-5637.

8. Zuo G H, Gu W, Fang H P, Zhou R H. Carbon Nanotube Wins the Competitive Binding over Proline-Rich Motif Ligand on SH3 Domain. J Phys Chem C 2011, 115: 12322-12328.

9. Ansari S A, Husain Q. Potential applications of enzymes immobilized on/in nano materials: A review. Biotechnol Adv 2012, 30: 512-523.

10. Coskun A, Banaszak M, Astumian R D, Stoddart J F, Grzybowski B A. Great expectations: can artificial molecular machines deliver on their promise? Chem Soc Rev 2012, 41: 19-30.

11. Guo X, Zhang J, Wang W. The interactions between nanomaterials and biomolecules and the microstructural features. Prog Phys 2012, 6: 285–293.

12. Shemetov A A, Nabiev I, Sukhanova A. Molecular interaction of proteins and peptides with nanoparticles. ACS nano 2012, 6: 4585-4602.

13. Walkey C D, Chan W C. Understanding and controlling the interaction of nanomaterials with proteins in a physiological environment. Chem Soc Rev 2012, 41: 2780-2799.

14. Mura S, Nicolas J, Couvreur P. Stimuli-responsive nanocarriers for drug delivery.

Nature materials 2013, 12: 991-1003.

15. Sapsford K E, Algar W R, Berti L, Gemmill K B, Casey B J, Oh E, Stewart M H, Medintz I L. Functionalizing nanoparticles with biological molecules: developing chemistries that facilitate nanotechnology. Chem Rev 2013, 113: 1904-2074.

16. Wei H, Wang E. Nanomaterials with enzyme-like characteristics (nanozymes): next-generation artificial enzymes. Chem Soc Rev 2013, 42: 6060-6093.

17. Dong S, Xiao H, Huang Q, Zhang J, Mao L, Gao S. Graphene Facilitated Removal of Labetalol in Laccase-ABTS System: Reaction Efficiency, Pathways and Mechanism. Sci Rep 2016, 6: 21396.

18. Cormack A N, Lewis R J, Goldstein A H. Computer simulation of protein adsorption to a material surface in aqueous solution: Biomaterials modeling of a ternary system. J Phys Chem B 2004, 108: 20408-20418.

19. Agashe M, Raut V, Stuart S J, Latour R A. Molecular simulation to characterize the adsorption behavior of a fibrinogen gamma-chain fragment. Langmuir 2005, 21: 1103-1117.

20. Kubiak K, Mulheran P A. Molecular dynamics simulations of hen egg white lysozyme adsorption at a charged solid surface. J Phys Chem B 2009, 113: 12189-12200.

21. Kuna J J, Voitchovsky K, Singh C, Jiang H, Mwenifumbo S, Ghorai PK, Stevens M M, Glotzer S C, Stellacci F. The effect of nanometre-scale structure on interfacial energy. Nature materials 2009, 8: 837-842.

22. Kokh D B, Corni S, Winn P J, Hoefling M, Gottschalk K E, Wade RC. ProMetCS: An Atomistic Force Field for Modeling Protein-Metal Surface Interactions in a

Continuum Aqueous Solvent. J Chem Theory Comput 2010, 6: 1753-1768.

23. Makarucha A J, Todorova N, Yarovsky I. Nanomaterials in biological environment: a review of computer modelling studies. Eur Biophys J 2011, 40: 103-115.

24. O'Brien E P, Straub J E, Brooks B R, Thirumalai D. Influence of Nanoparticle Size and Shape on Oligomer Formation of an Amyloidogenic Peptide. J Phys Chem Lett 2011, 2: 1171-1177.

25. Vila Verde A, Beltramo P J, Maranas J K. Adsorption of homopolypeptides on gold investigated using atomistic molecular dynamics. Langmuir 2011, 27: 5918-5926.

26. Zuo G, Zhou X, Huang Q, Fang H P, Zhou R H. Adsorption of Villin Headpiece onto Graphene, Carbon Nanotube, and C60: Effect of Contacting Surface Curvatures on Binding Affinity. J Phys Chem C 2011, 115: 23323-23328.

27. Brancolini G, Kokh D B, Calzolai L, Wade R C, Corni S. Docking of ubiquitin to gold nanoparticles. ACS nano 2012, 6: 9863-9878.

28. Deighan M, Pfaendtner J. Exhaustively sampling peptide adsorption with metadynamics. Langmuir 2013, 29: 7999-8009.

29. Ding F, Radic S, Chen R, Chen P, Geitner N K, Brown J M, Ke P C. Direct observation of a single nanoparticle-ubiquitin corona formation. Nanoscale 2013, 5: 9162-9169.

30. Kang S G, Huynh T, Xia Z, Zhang Y, Fang H, Wei G, Zhou R. Hydrophobic interaction drives surface-assisted epitaxial assembly of amyloid-like peptides. J Am Chem Soc 2013, 135: 3150-3157.

31. Li R, Chen R, Chen P, Wen Y, Ke P C, Cho S S. Computational and experimental

characterizations of silver nanoparticle-apolipoprotein biocorona. J Phys Chem B 2013, 117: 13451-13456.

32. Bhirde A A, Hassan S A, Harr E, Chen X. Role of Albumin in the Formation and Stabilization of Nanoparticle Aggregates in Serum Studied by Continuous Photon Correlation Spectroscopy and Multiscale Computer Simulations. J Phys Chem C Nanomater Interfaces 2014, 118: 16199-16208.

33. Nawrocki G, Cieplak M. Aqueous Amino Acids and Proteins Near the Surface of Gold in Hydrophilic and Hydrophobic Force Fields. J Phys Chem C 2014, 118: 12929-12943.

34. Pan H, Qin M, Meng W, Cao Y, Wang W. How do proteins unfold upon adsorption on nanoparticle surfaces? Langmuir 2012, 28: 12779-12787.

35. Laio A, Parrinello M. Escaping free-energy minima. Proc Natl Acad Sci U S A 2002, 99: 12562-12566.

36. Alessandro L, Francesco L G. Metadynamics: a method to simulate rare events and reconstruct the free energy in biophysics, chemistry and material science. Rep Prog Phys 2008, 71: 126601.

37. Bonomi M, Branduardi D, Bussi G, Camilloni C, Provasi D, Raiteri P, Donadio D, Marinelli F, Pietrucci F, Broglia R A, Parrinello M. PLUMED: A portable plugin for free-energy calculations with molecular dynamics. Comput Phys Commun 2009, 180: 1961-1972.

38. Tribello G A, Bonomi M, Branduardi D, Camilloni C, Bussi G. PLUMED 2: New feathers for an old bird. Comput Phys Commun 2014, 185: 604-613.

39. Tadros, T. Encyclopedia of Colloid and Interface Science. Wokingham: Springer Link: Berlin, Heidelberg, 2013. 1436p.

40. Wang B, Merz K M. A Fast QM/MM (Quantum Mechanical/Molecular Mechanical) Approach to Calculate Nuclear Magnetic Resonance Chemical Shifts for Macromolecules. J Chem Theory Comput 2006, 2: 209-215.

41. Lindorff-Larsen K, Piana S, Palmo K, Maragakis P, Klepeis J L, Dror R O, Shaw D E. Improved side-chain torsion potentials for the Amber ff99SB protein force field. Proteins 2010, 78: 1950-1958.

42. Berendsen H J C, van der Spoel D, van Drunen R. GROMACS: A message-passing parallel molecular dynamics implementation. Comput Phys Commun 1995, 91: 43-56.

43. Hess B, Kutzner C, van der Spoel D, Lindahl E. GROMACS 4: Algorithms for Highly Efficient, Load-Balanced, and Scalable Molecular Simulation. J Chem Theory Comput 2008, 4: 435-447.

44. Tiwary P, Parrinello M. A time-independent free energy estimator for metadynamics. J Phys Chem B 2015, 119: 736-742.

45. Daura X, Gademann K, Jaun B, Seebach D, van Gunsteren W F, Mark A E. Peptide folding: When simulation meets experiment. Angew Chem Int Edit 1999, 38: 236-240.

46. Kabsch W, Sander C. Dictionary of protein secondary structure: pattern recognition of hydrogen-bonded and geometrical features. Biopolymers 1983, 22: 2577-2637.

47. Touw W G, Baakman C, Black J, te Beek TA, Krieger E, Joosten R P, Vriend G. A series of PDB-related databanks for everyday needs. Nucleic Acids Res 2015, 43: D364-368.

48. Somavarapu A K, Kepp K P. The Dependence of Amyloid-beta Dynamics on Protein Force Fields and Water Models. Chem phys chem 2015, 16: 3278-3289.

49. Carballo-Pacheco M, Strodel B. Comparison of force fields for Alzheimer's A 42: A case study for intrinsically disordered proteins. Protein Sci 2017, 26: 174-185.

50. Bowman G R. Accurately Modeling Nanosecond Protein Dynamics Requires at least Microseconds of Simulation. J Comput Chem 2016, 37: 558-566.

51. Berhanu W M, Hansmann U H E. Side-chain hydrophobicity and the stability of A beta(16-22) aggregates. Protein Sci 2012, 21: 1837-1848.

52. Smith M D, Rao J S, Segelken E, Cruz L. Force-Field Induced Bias in the Structure of A beta(21-30): A Comparison of OPLS, AMBER, CHARMM, and GROMOS Force Fields. J Chem Inf Model 2015, 55: 2587-2595.

53. Rosenman D J, Wang C Y, Garcia A E. Characterization of A beta Monomers through the Convergence of Ensemble Properties among Simulations with Multiple Force Fields. J Phys Chem B 2016, 120: 259-277.

54. Petrov D, Zagrovic B. Are Current Atomistic Force Fields Accurate Enough to Study Proteins in Crowded Environments? PLoS Comp Biol 2014, 10: e1003638.

55. Gao Y, Zhang C M, Zhang J Z H, Mei Y. Evaluation of the Coupled Two-Dimensional Main Chain Torsional Potential in Modeling Intrinsically Disordered Proteins. J Chem Inf Model 2017, 57: 267-274.

**Figure Captions**

**Figure 1.** The model system for studying the adsorption of GB1 upon the polystyrenes (PS) surface. The protein is colored in blue and cyan. The blue spheres represent the residues in the hydrophobic core. The PS surface is colored gray. The red spheres are the Cl- ions restrained on the surface, in order to mimic the electrical double layer effect.

**Figure 2.** The two-dimensional FELs calculated from metadynamics simulations. Each corresponds to a different simulation. The energy scale is shown on the top right and the unit is kJ/mol. Three major proteins states are labeled as the folded (F), the intermediate (I) and the unfolded (U) states, respectively. Each protein state involves both the adsorbed and desorbed states. Six basins of attraction are identified in (a) and labelled from I to VI, respectively. The pink lines are the superimposed trajectories. The arrows indicate the transition events between the states. The color of the arrows is explained in the text.

**Figure 3.** The evolution of the secondary structures as a function of time. From top down, each figure corresponds to a trajectory in Fig. 2.

**Figure 4.** The mapping of the vdW and electrostatic energies between protein and surface onto residues, averaged on the conformations collected from the respective state. The three rows correspond to the F-, I- and U-states, respectively. The protein sequence, the color code for secondary structure in the native state, and the color code for the hydrophobicity and electrical properties of residues are given on the top.

**Figure 5.** The central structures of the largest clusters, corresponding to the basins from I to VI shown in Fig. 2(a), respectively. The belonged basin of each structure is marked on the top-right corner. The residues contributing the most to the vdW energy are colored violet, while that contributing the most to the electrostatic energy are colored red.

**Figure 6.** (a) The scenario suggested by the experiment34. Double ended arrows indicate that the reactions are reversible. (b) The scenario deduced from our simulations. The solid lines indicate the reactions that were observed in the simulations, while the dash lines indicate the reactions that were not, possibly due to their low probabilities in an unfavorable environment and finite simulation time.

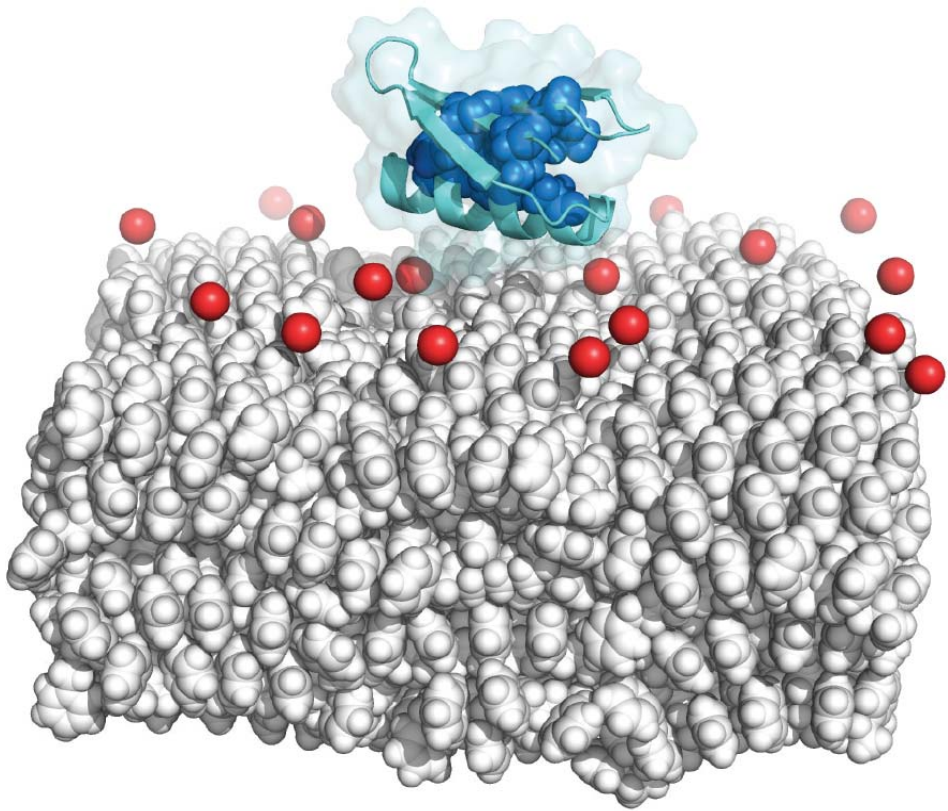

**Figure 1.** The model system for studying the adsorption of GB1 upon the polystyrenes (PS) surface. The protein is colored in blue and cyan. The blue spheres represent the residues in the hydrophobic core. The PS surface is colored gray. The red spheres are the Cl- ions restrained on the surface, in order to mimic the electrical double layer effect.

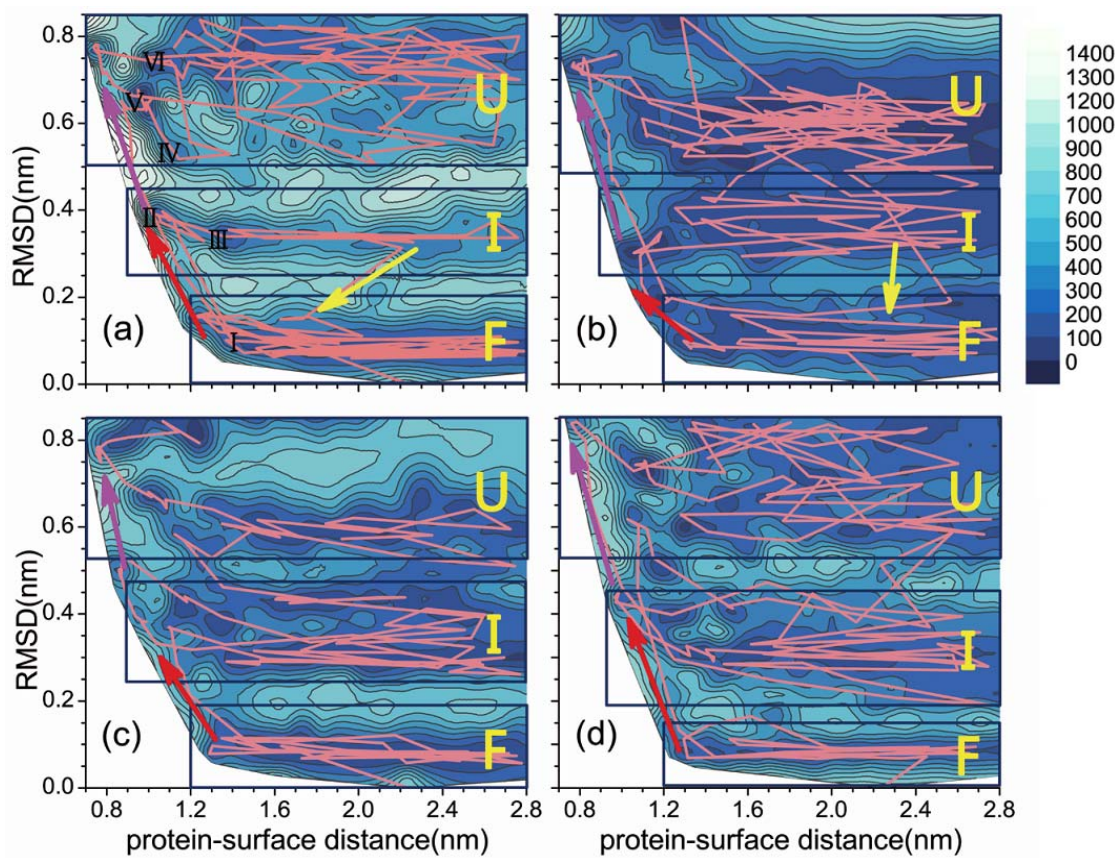

**Figure 2.** The two-dimensional FELs calculated from metadynamics simulations. Each corresponds to a different simulation. The energy scale is shown on the top right and the unit is kJ/mol. Three major proteins states are labeled as the folded (F), the intermediate (I) and the unfolded (U) states, respectively. Each protein state involves both the adsorbed and desorbed states. Six basins of attraction are identified in (a) and labelled from I to VI, respectively. The pink lines are the superimposed trajectories. The arrows indicate the transition events between the states. The color of the arrows is explained in the text.

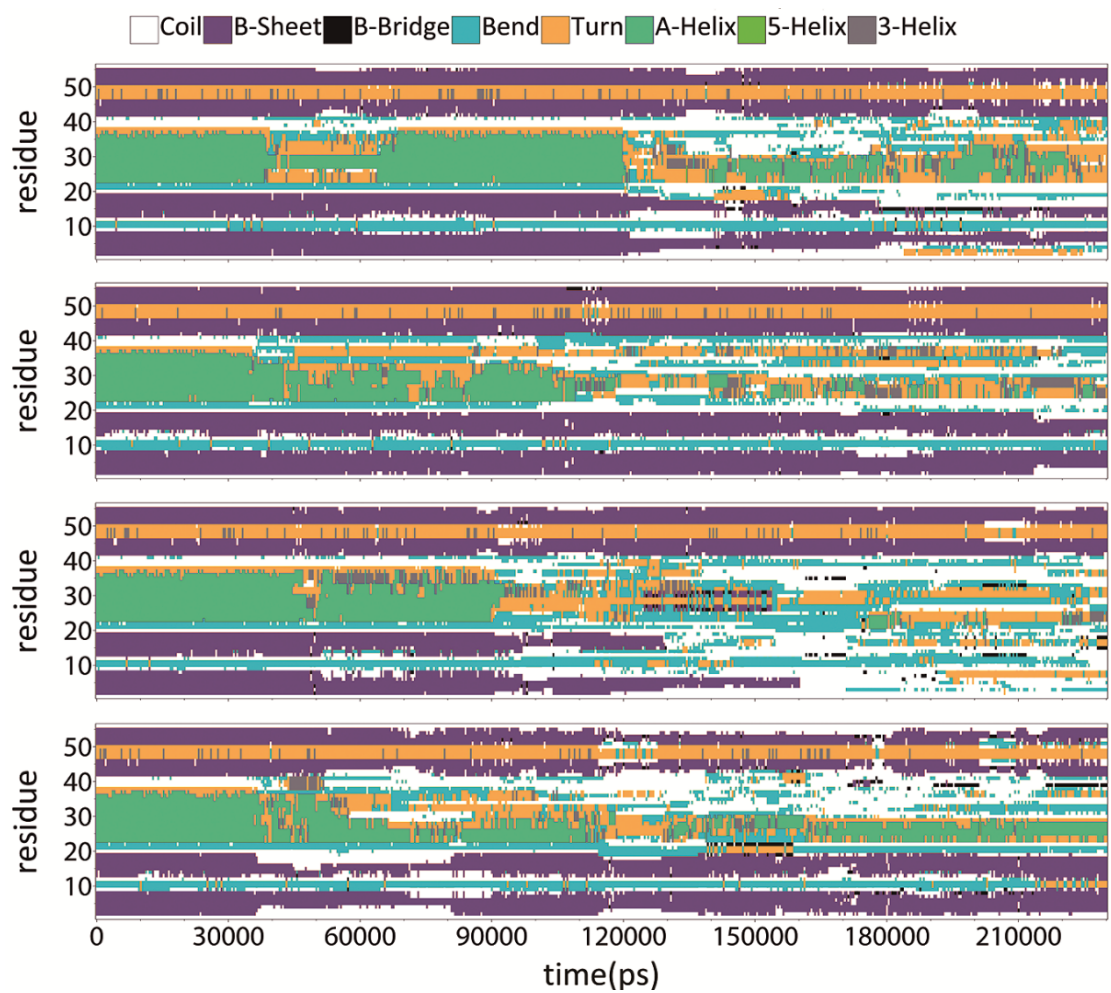

**Figure 3.** The evolution of the secondary structures as a function of time. From top down, each figure corresponds to a trajectory in Fig. 2.

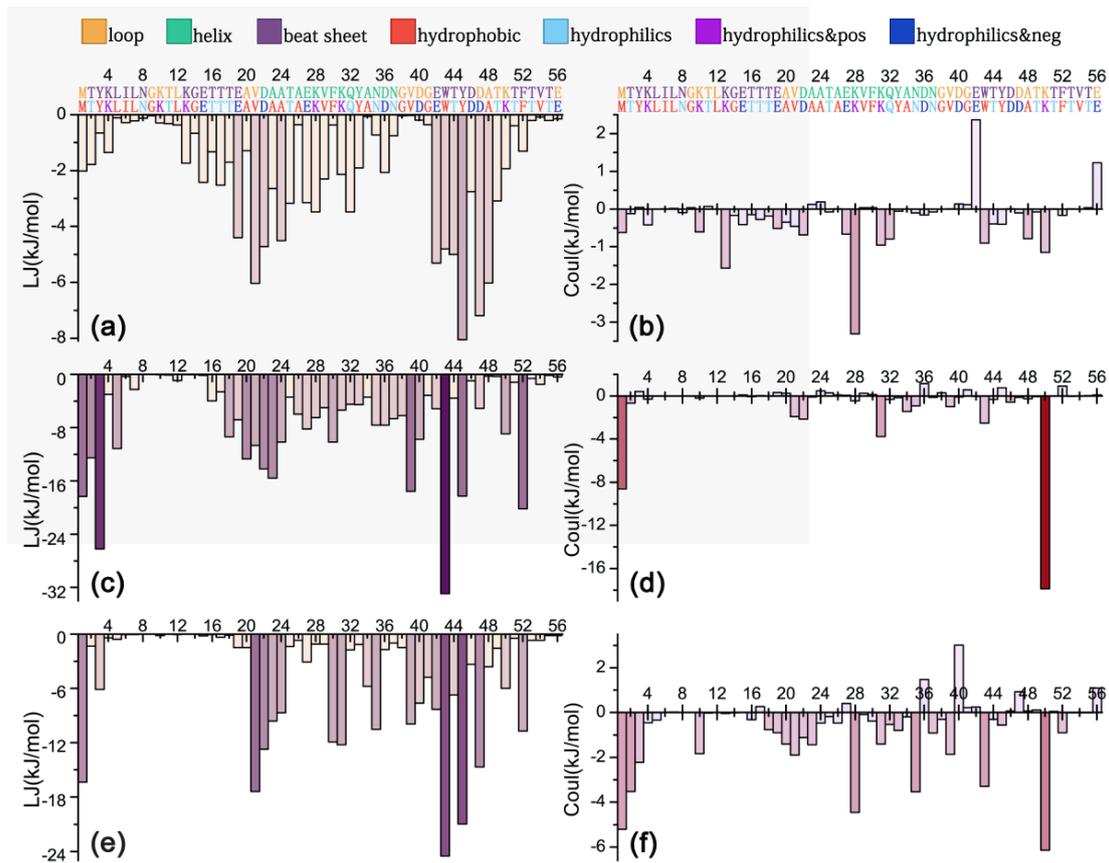

**Figure 4.** The mapping of the vdW and electrostatic energies between protein and surface onto residues, averaged on the conformations collected from the respective state. The three rows correspond to the F-, I- and U-states, respectively. The protein sequence, the color code for secondary structure in the native state, and the color code for the hydrophobicity and electrical properties of residues are given on the top.

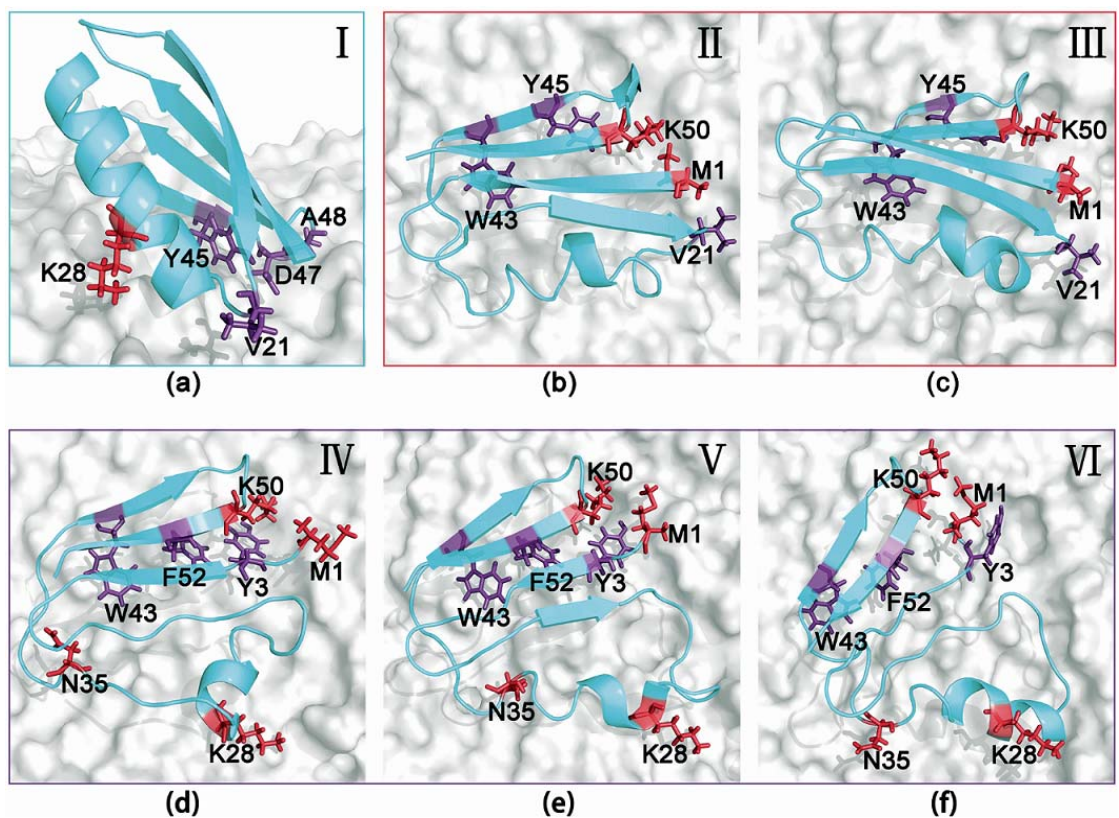

**Figure 5.** The central structures of the largest clusters, corresponding to the basins from I to VI shown in Fig. 2(a), respectively. The belonged basin of each structure is marked on the top-right corner. The residues contributing the most to the vdW energy are colored violet, while that contributing the most to the electrostatic energy are colored red.

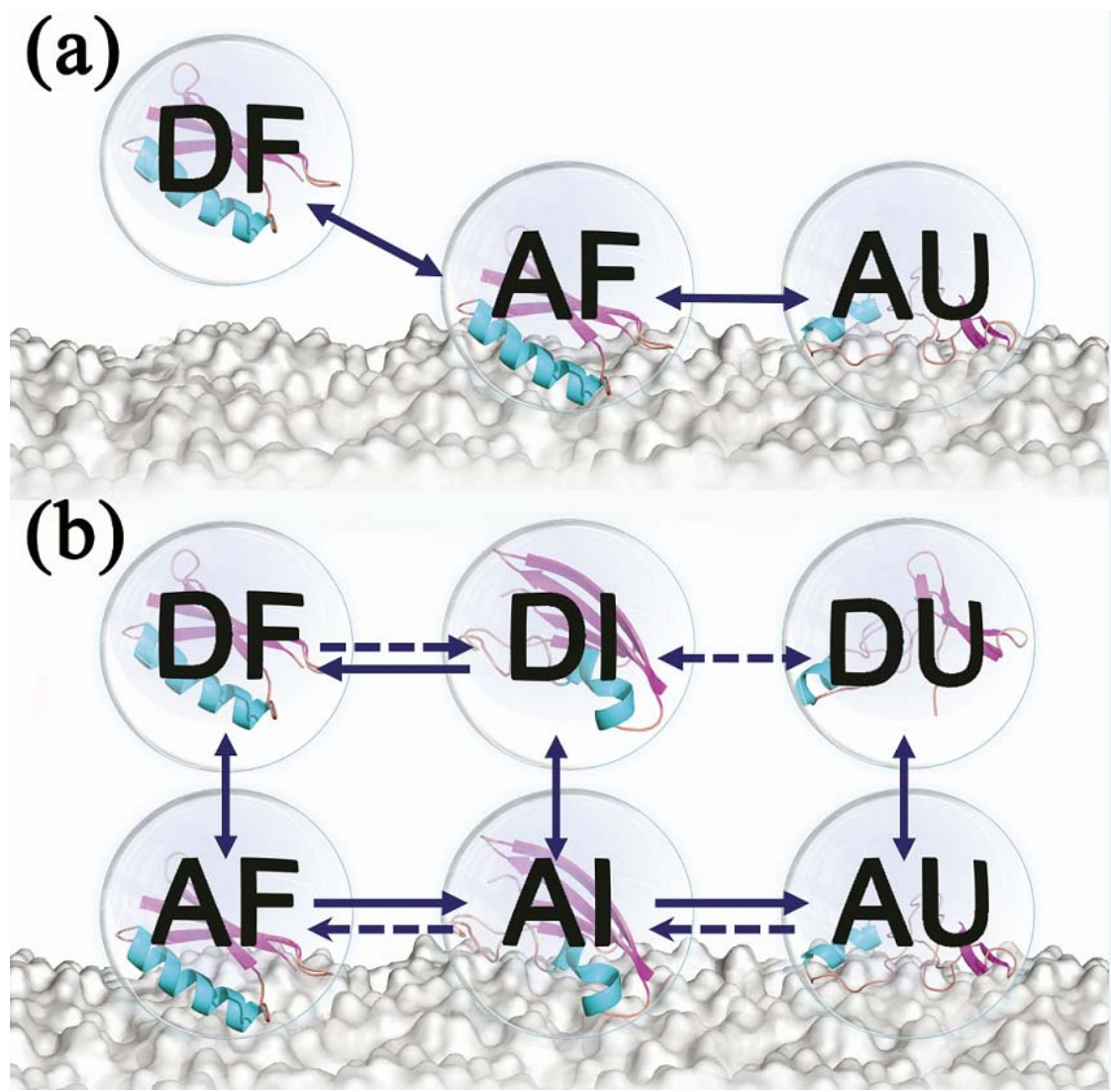

**Figure 6.** (a) The scenario suggested by the experiment[34]. Double ended arrows indicate that the reactions are reversible. (b) The scenario deduced from our simulations. The solid lines indicate the reactions that were observed in the simulations, while the dash lines indicate the reactions that were not, possibly due to their low probabilities in an unfavorable environment and finite simulation time.